\documentclass[showpacs,10pt,twocolumn,prl]{revtex4-1}
\usepackage{amsmath}
\usepackage{amssymb}
\usepackage{graphics}
\usepackage{epsfig}
\usepackage{color}

\begin{document}


\title{Structures and physical properties of V-based kagome metals CsV$_{6}$Sb$_{6}$ and CsV$_{8}$Sb$_{12}$}
\author{Qiangwei Yin, Zhijun Tu, Chunsheng Gong, Shangjie Tian, and Hechang Lei$^{*}$}
\affiliation{Department of Physics and Beijing Key Laboratory of Opto-electronic Functional Materials $\&$ Micro-nano Devices, Renmin University of China, Beijing 100872, China
}
\date{\today}

\begin{abstract}

We report two new members of V-based kagome metals CsV$_{6}$Sb$_{6}$ and CsV$_{8}$Sb$_{12}$. The most striking structural feature of CsV$_{6}$Sb$_{6}$ is the V kagome bilayers. For CsV$_{8}$Sb$_{12}$, there is an intergrowth of two-dimensional  V kagome layers and one-dimensional V chains and the latter lead to the orthorhombic symmetry of this material.
Further measurements indicate that these two materials exhibit metallic and Pauli paramagnetic behaviors. More importantly, different from CsV$_{3}$Sb$_{5}$, the charge density wave state and superconductivity do not emerge in CsV$_{6}$Sb$_{6}$ and CsV$_{8}$Sb$_{12}$ when temperature is above 2 K.
Small magnetoresistance with saturation behavior and linear field dependence of Hall resistivity at high field and low temperature suggest that the carriers in both materials should be uncompensated with much different concentrations.
The discovery of these two new V-based kagome metals sheds light on the exploration of correlated topological materials based on kagome lattice.

\end{abstract}


\maketitle

\section{Introduction}

Two-dimensional (2D) kagome lattice is a paradigm to study the effects of strongly geometrical frustration which could host many exotic magnetic ground states like quantum spin liquid state with fractionalized excitations \cite{Balents,HanTH}.
When introducing charge degrees of freedom, Janus-like kagome lattice exhibits nontrivial topological electronic structures with unusual features of Dirac nodal points, flat band and saddle point \cite{YeL,LiuZ,LinZ,KangM}.
More importantly, the combination of magnetic correlation and band topology leads to the emergence of various of exotic correlated topological phenomena in metallic materials with kagome lattice (kagome metals), such as large anomalous Hall effect (AHE) \cite{Nakatsuji,YeL,LiuE,WangQ}, negative magnetism of flat band \cite{YinJX}, large magnetic-field tunability \cite{YinJX2} and the formation of Chern gap with edge state \cite{YinJX3}.

Very recently, the kagome metals with electron correlations other than magnetic one have been extended to the V-based materials AV$_{3}$Sb$_{5}$ (A = K, Rb and Cs) and they exhibit the coexistence of charge density wave (CDW) state and superconductivity with a non-zero $Z_{2}$ topological invariant \cite{Ortiz1,Ortiz2,Ortiz3,YinQW}.
Such phenomena are closely related to the physics of van Hove filling in kagome lattice \cite{YuSL,WangWS,Kiesel}.
Moreover, the CDW state with three-dimensional (3D) 2$\times$2$\times$2 superlattice \cite{LiangZ,LiHX} shows some exotic behaviors.
For instance, the CDW state may have a chirality which could result in large anomalous Hall conductivity without long-range ferromagnetism \cite{JiangYX,YangSY,YuFH,FengX}.
In addition, there is an intricate relationship between CDW state and superconductivity.
For example, when the CDW transition is suppressed with pressure monotonically, the superconductivity shows an unusual multiple-dome feature with a significant enhancement \cite{ChenKY,YuFH2,ChenX,ZhaoCC,DuF}. The roton pair density wave state has also been observed in the superconducting state \cite{ChenH}.

In order to understand these diverse properties of V-based kagome metals further, not only the comprehensive studies on AV$_{3}$Sb$_{5}$ but also the exploration of novel V-based kagome materials is important.
For many of magnetic kagome metals, a series of compounds with the key kagome layer exist, such as the binary T$_{m}$X$_{n}$ kagome metals (T = Mn, Fe, Co; X = Sn, Ge; $m$ : $n$ = 3 : 1, 3 : 2, 1 : 1) \cite{KangM}.
Systematic studies on these materials can provide some important clues to understand the effects of different local structural environment and strength of interlayer coupling on the properties of kagome layer.
But such kind of homologous compounds is still scarce in V-based kagome materials.
In this work, we report the discovery of V-based kagome metals CsV$_{6}$Sb$_{6}$ and CsV$_{8}$Sb$_{12}$, two homologous compounds of CsV$_{3}$Sb$_{5}$. The characterizations of physical properties of single crystals indicate that these compounds show metallic behaviors without CDW and superconducting transitions down to 2 K.

\section{Methods}

Single crystals of CsV$_{6}$Sb$_{6}$ and CsV$_{8}$Sb$_{12}$ were grown using self-flux method.
High-purity Cs (ingot), V (powder) and Sb (grain) were taken in a stoichiometric molar ratio of 1 : 6 : 30 for CsV$_{6}$Sb$_{6}$ and 1 : 8 : 50 for CsV$_{8}$Sb$_{12}$, respectively, and placed in a quartz tube with alumina crucible under partial argon atmosphere.	
The sealed quartz tubes for both materials were heated up to 1373 K for 12h and soaked there for another 24 h.
After that, for CsV$_{6}$Sb$_{6}$, the temperature was rapidly cooled down to 1273 K with subsequently cooling down to 1163 K at 1.5 K/h, and then the temperature was held there for 70 h.
For CsV$_{8}$Sb$_{12}$, the temperature was cooled down directly to 1123 K with the rate of 2 K/h.
Finally, the ampoules were taken out of furnace and the single crystals were separated from the flux by a centrifuge.
Shiny crystals with typical sizes of 1$\times$1$\times$0.02 mm$^{3}$ (length$\times$width$\times$thickness) for CsV$_{6}$Sb$_{6}$ and 4$\times$1$\times$0.1 mm$^{3}$ for CsV$_{8}$Sb$_{12}$ can be obtained. Both crystals are stable in air.
In order to prevent the reaction of Cs with air and water, all the preparation processes except the sealing and heat treatment procedures are carried out in an argon-filled glove box.
The elemental analysis was performed using the energy-dispersive X-ray spectroscopy (EDX).
XRD patterns were collected using a Bruker D8 X-ray diffractometer with Cu $K_{\alpha}$ radiation ($\lambda=$ 1.5418 \AA) at room temperature.
Single crystal XRD patterns at 300 K were collected using a Bruker D8 VENTURE PHOTO II diffractometer with multilayer mirror monochromatized Mo $K\alpha$ ($\lambda=$ 0.71073 \AA) radiation. Unit cell refinement and data merging were done with the SAINT program, and an absorption correction was applied using Multi-Scans.
The structural solutions were obtained by intrinsic phasing methods using the program APEX3 \cite{APEX}, and the final refinement was completed with the SHELXL suite of programs \cite{Sheldrick}.
Electrical transport measurements were carried out in a Quantum Design physical property measurement system (PPMS-14T). The field dependence of $ab$-plane longitudinal and transverse electrical resistivity were measured using a five-probe method and the current flows in the $ab$ plane of the crystal.
The magnetoresistance and Hall resistivity were obtained from symmetrizing and antisymmetrizing the longitudinal and transverse resistivity measured at the positive and negative fields, respectively.
The $c$-axis resistivity was measured by attaching current and voltage wires on the opposite sides of the plate-like crystal.
Magnetization measurements were performed in a Quantum Design magnetic property measurement system (MPMS3).

\section{Results and Discussion}

\begin{figure}[tbp]
\centerline{\includegraphics[scale=0.46]{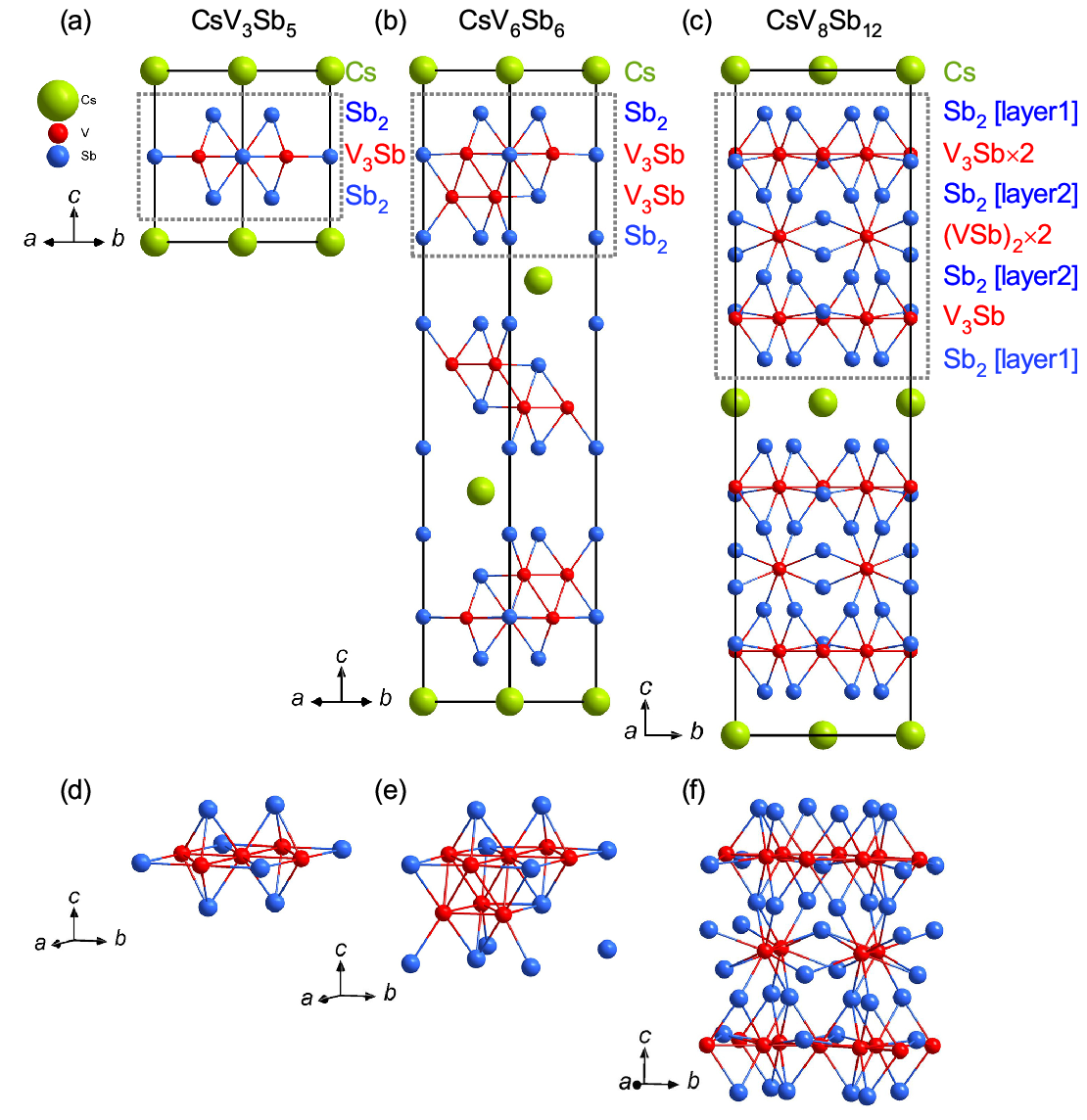}} \vspace*{-0.3cm}
\caption{Front view of structures of (a) CsV$_{3}$Sb$_{5}$, (b) CsV$_{6}$Sb$_{6}$, and (c) CsV$_{8}$Sb$_{12}$. The big green, small red and medium blue balls represent Cs, V, and Sb atoms. The black lines represent the unit cell edges. (d) - (f) Side view of V-Sb building blocks emphasized by dotted rectangles in (a) - (c).}
\end{figure}

\begin{figure}[tbp]
\centerline{\includegraphics[scale=0.46]{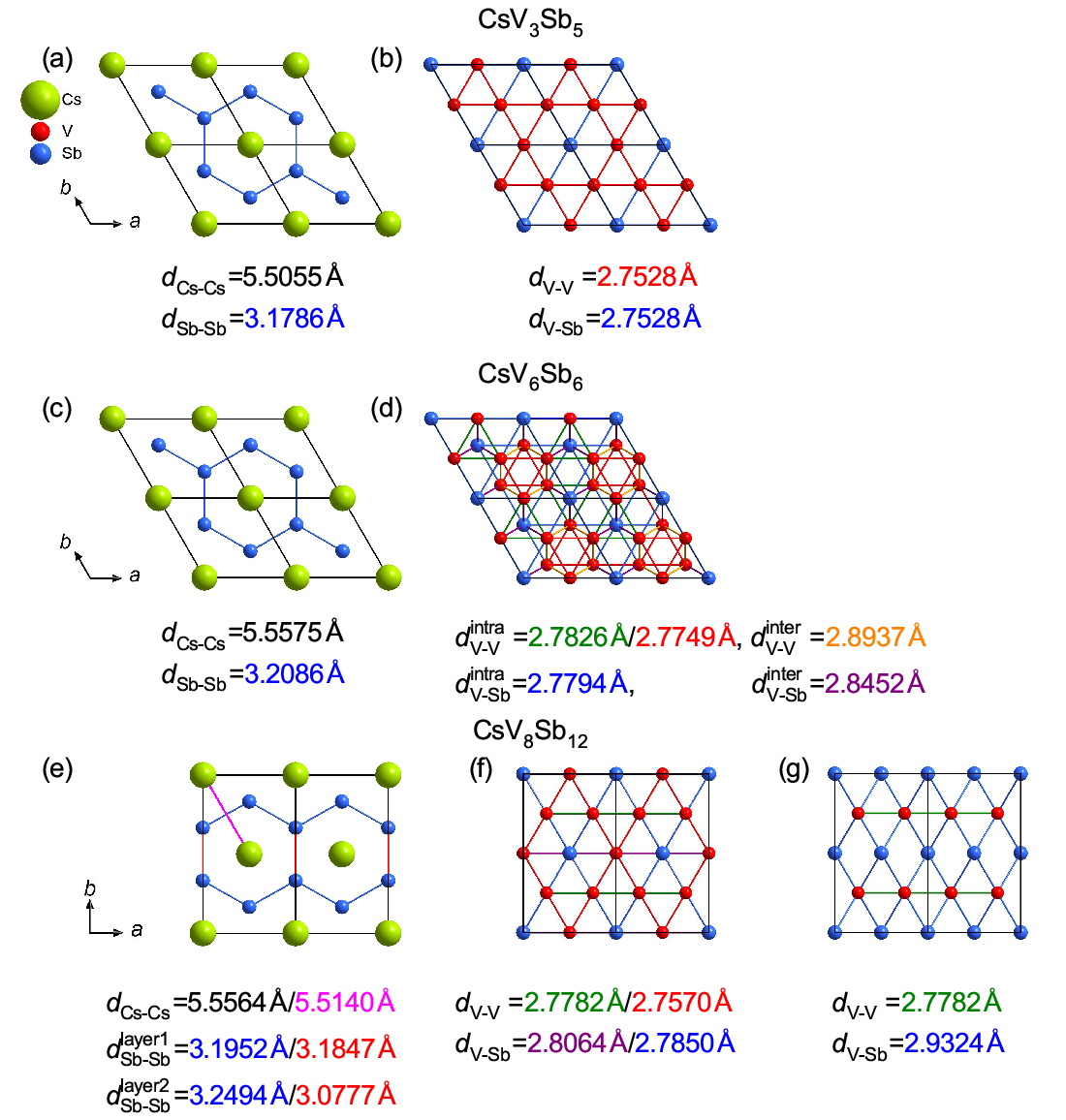}} \vspace*{-0.3cm}
\caption{Top view of (a) Cs, Sb$_{2}$ layers and (b) V$_{3}$Sb layer of CsV$_{3}$Sb$_{5}$. (c) Cs, Sb$_{2}$ layers and (d) 2V$_{3}$Sb bilayer of CsV$_{6}$Sb$_{6}$. (e) Cs, Sb$_{2}$ layers, (f) V$_{3}$Sb layer, and (g) (VSb)$_{2}$ layer of CsV$_{8}$Sb$_{12}$. The corresponding Cs-Cs, Sb-Sb, V-V, V-Sb atomic distances in each layer are shown below the figures.}
\end{figure}

As shown in Fig. 1(a), CsV$_{3}$Sb$_{5}$ has a layered structure with hexagonal symmetry (space group $P6/mmm$, No. 191). The $a$- and $c$-axial lattice parameters are 5.5055 \AA\  and 9.3287 \AA\ \cite{Ortiz1}. It consists of Cs, Sb and V-Sb layer stacking along $c$ axis alternatively (Fig. 1(a)) \cite{Ortiz1}. For the Cs layer, Cs atoms form a perfect triangle lattice with the distance of Cs atoms $d_{\rm Cs-Cs}=$ 5.5055 \AA\ (Fig. 2(a)). For the Sb layer, Sb atoms compose a honeycomb lattice with $d_{\rm Sb-Sb}=$ 3.1786 \AA\ and Cs atoms locate at the centers of each hexagons (Fig. 2(a)). Because there are two Sb atoms in one unit cell, this layer is denoted by the Sb$_{2}$ layer. The key structural ingredient of CsV$_{3}$Sb$_{5}$ is the 2D undistorted V kagome lattice in the V-Sb layer and the distance of V atoms $d_{\rm V-V}=$ 2.7528 \AA\ (Fig. 2(b)). The Sb atoms occupy at the centers of V hexagons with $d_{\rm V-Sb}=$ 2.7528 \AA\ when Sb atoms in the Sb$_{2}$ layer locate below and above the centers of V triangles (Fig. 1(d)). This V-Sb layer can be denoted by V$_{3}$Sb layer due to three V and one Sb atoms in one unit cell. Because there are one Cs, two Sb$_{2}$ and one V$_{3}$Sb layers in each unit cell, the chemical formula of CsV$_{3}$Sb$_{5}$ can be expressed as CsV$_{3}$Sb$_{5}$ = Cs + 2Sb$_{2}$ + V$_{3}$Sb.

For layered CsV$_{6}$Sb$_{6}$, it has a similar hexagonal structural symmetry to CsV$_{3}$Sb$_{5}$ (space group $R\bar{3}m$, No. 166) with $a=$ 5.5575(8) \AA\ and $c=$ 35.165(9) \AA\ (Table 1). The structures of Cs layer and Sb$_{2}$ layer with Sb1 and Sb2 sites in CsV$_{6}$Sb$_{6}$ are also similar to those in CsV$_{3}$Sb$_{5}$ and the former has slightly larger $d_{\rm Cs-Cs}$ (= 5.5055 \AA) and $d_{\rm Sb-Sb}$ (= 3.2086 \AA) (Figs. 1(b) and 2(c)).
Instead of the V$_{3}$Sb monolayers in CsV$_{3}$Sb$_{5}$, however, there are 2V$_{3}$Sb bilayers with V and Sb1 sites in CsV$_{6}$Sb$_{6}$ (Figs. 1(b) and 2(d)), similar to that in Fe$_{3}$Sn$_{2}$ \cite{WangQ1}.
The shift of top V$_{3}$Sb layer along the direction of 1/3$a$ - 1/3$b$ will coincide with the bottom one. Due to this shift, the Sb atoms in one layer locate below and above the centers of V triangles in another layer when another half of V triangles in both layers form the distorted octahedra of V atoms (Figs. 1(e) and 2(d)).
Such inequivalent local environments of V triangles also lead to the distortion of kagome layers with two kinds of equilateral triangles which have different intralayer V-V distances $d_{\rm V-V}^{\rm intra}=$ 2.7826 \AA\ (green ones) and 2.7749 \AA\ (red ones). The former one with smaller value could be ascribed to the stronger V-V interactions in the V octahedra. Because of this distortion, the intralayer V-Sb distance $d_{\rm V-Sb}^{\rm intra}$ (= 2.7794 \AA) falls in between two values of $d_{\rm V-V}^{\rm intra}$.
In contrast, the interlayer V-V and V-Sb distances ($d_{\rm V-V}^{\rm inter}=$ 2.8937 \AA\ and $d_{\rm V-Sb}^{\rm inter}=$ 2.8452 \AA) are much larger than those of $d_{\rm V-V}^{\rm intra}$ and $d_{\rm V-Sb}^{\rm intra}$, reflecting the relatively weak interlayer interaction when compared to the intralayer one. It has to be noted that the Sb atoms in the kagome layer are slightly move along the $c$ axis and toward to the center of bilayer (see the different $z$ values of V and Sb1 in Table 2). Since there are three Cs, six Sb$_{2}$ and three V$_{3}$Sb bilayers in one unit cell, we have 3CsV$_{6}$Sb$_{6}$ = 3Cs + 6Sb$_{2}$ + 3$\times$2V$_{3}$Sb.

CsV$_{8}$Sb$_{12}$ has an orthorhombic symmetry (space group $Fmmm$, No. 69) with $a=$ 5.5564(3) \AA, $b=$ 9.5260(5) \AA, and $c=$ 36.227(2) \AA\ (Table 1). Although the structure of CsV$_{8}$Sb$_{12}$ is much more complicated than CsV$_{3}$Sb$_{5}$ and CsV$_{6}$Sb$_{6}$, these three materials still share some common structure features.
The local environments of Cs layer, Sb$_{2}$ layer with Sb2 and Sb3 sites and V$_{3}$Sb layer with V1, V2, and Sb1 sites in CsV$_{8}$Sb$_{12}$ (Figs. 1(c), 2(e) and 2(f)) are similar to those in CsV$_{3}$Sb$_{5}$ and CsV$_{6}$Sb$_{6}$.
But they are compressed along the $b$ axis of orthorhombic lattice (the [210] direction in hexagonal lattice) and it leads to the distortions of these layers with inequivalent intralayer atomic distances, such as $d_{\rm Cs-Cs}$ = 5.5564/5.5140 \AA, $d_{\rm V-V}$ = 2.7782/2.7570 \AA, and $d_{\rm V-Sb}$ = 2.8064/2.7850 \AA\ (Figs. 2(e) and 2(f)).
In addition, such compression also results in the significant movement of Sb atoms in the kagome layer along the $c$ axis (Figs. 1(c) and 1(f) as well as Table 2). Correspondingly, the $d_{\rm Sb-Sb}$ in both Sb$_{2}$ layers above and below the V kagome layer (labelled as Sb$_{2}$ layer 1 and layer 2 in Fig. 1(c)) become different. The $d_{\rm Sb-Sb}$ in Sb$_{2}$ layer 1 are 3.1952 \AA\ and 3.1847 \AA\ when those in Sb$_{2}$ layer 2 are 3.2494 \AA\ and 3.0777 \AA\ (Fig. 2(e)).
The most distinctive structural feature of CsV$_{8}$Sb$_{12}$ is the (VSb)$_{2}$ layer with V3 and Sb4 sites between two V$_{3}$Sb kagome layers (Figs. 1(c) and 2(g)). In this layer, there are two Sb layers locating below and above a V layer with $d_{\rm V-Sb}$ = 2.9324 \AA. For the former one, the in-plane arrangement of Sb atoms is similar to that in the V$_{3}$Sb layer (Fig. 1(f)). For the later one, V atoms form one dimensional (1D) chains along the $a$ direction with $d_{\rm V-V}$ = 2.7782 \AA\ (Figs. 1(f) and 2(g)), which is exactly same as the value of $d_{\rm V-V}$ in the V$_{3}$Sb layer along the $a$ axis. Actually, such 1D chains of V atoms should lead to the orthorhombic symmetry of CsV$_{8}$Sb$_{12}$.
There are two Cs, eight Sb$_{2}$, four V$_{3}$Sb, and two VSb$_{2}$ layers in one unit cell and each layer contains twice atoms when compared to CsV$_{3}$Sb$_{5}$ and CsV$_{6}$Sb$_{6}$, thus we have 4CsV$_{8}$Sb$_{12}$ = 2$\times$[2Cs + 8Sb$_{2}$ + 4V$_{3}$Sb + 2(VSb)$_{2}$].

\begin{table}
\centering
\caption{Crystallographic data of CsV$_{6}$Sb$_{6}$ and CsV$_{8}$Sb$_{12}$ at 300 K.}
\begin{tabular}{ccc}
\hline\hline
chemical formula                          & CsV$_{6}$Sb$_{6}$        & CsV$_{8}$Sb$_{12}$        \\
space group                               & $R\bar{3}m$              & $Fmmm$                    \\
crystal system                            & rhombohedral             & orthorhombic              \\
$a$ ({\AA})                               & 5.5575(8)                & 5.5564(3)                 \\
$b$ ({\AA})                               & 5.5575(8)                & 9.5260(5)                 \\
$c$ ({\AA})                               & 35.165(9)                & 36.227(2)                 \\
$V$ ({\AA}$^{3}$)                         & 940.59(36)               & 1917.50(18)               \\
$Z$                                       & 3                        & 4                         \\
dimens                                    & 0.05/0.40/0.50           & 0.10/0.36/0.39            \\
min/mid/max(mm$^{3}$)                     &                          &                           \\
calcd density (g cm$^{-3}$)               & 6.047                    & 6.880                     \\
abs coeff (mm$^{-1}$)                     & 19.270                   & 21.971                    \\
$h$                                       & -7 $\leq$ $h$ $\leq$ 7   & -7 $\leq$ $h$ $\leq$ 6    \\
$k$                                       & -6 $\leq$ $k$ $\leq$ 6   & -12 $\leq$ $k$ $\leq$ 11  \\
$l$                                       & -46 $\leq$ $l$ $\leq$ 46 & -46 x$\leq$ $l$ $\leq$ 48 \\
reflns                                    & 2536/334/0.0642          & 5846/709/0.0697           \\
collected/unique/$R$(int)                 &                          &                           \\
data/params/restraints                    & 334/18/0                 & 709/36/0                  \\
GOF on $F^{2}$                            & 1.139                    & 1.139                     \\
$R$ indices (all data)                    & 0.0825/0.2643            & 0.0590/0.1824             \\
($R$1/$wR$2)$^{a}$                        &                          &                           \\
\hline\hline
\end{tabular}
\label{1}
\end{table}

\begin{table}
\centering
\caption{Atomic positions, s.o.f. and equivalent isotropic displacement parameters $U_{\rm eq}$ obtained from the XRD fits for CsV$_{6}$Sb$_{6}$ and CsV$_{8}$Sb$_{12}$ at 300 K.}
\begin{tabular}{ccccccc}
\hline\hline
\multicolumn{7}{c}{CsV$_{6}$Sb$_{6}$}                                                   \\
atom & site  & $x/a$      & $y/b$      & $z/c$       & s.o.f. & $U_{\rm eq}$ (A$^{2}$)  \\
Cs   & 3$a$  & 2/3        & 1/3        & 1/3         & 1      & 0.0390(11)              \\
V    & 18$h$ & 0.6662(4)  & 83331(2)   & 0.20093(7)  & 1      & 0.0146(10)              \\
Sb1  & 6$c$  & 2/3        & 1/3        & 0.19918(5)  & 0.9548 & 0.0148(9)               \\
Sb2  & 6$c$  & 1/3        & 2/3        & 0.26503(5)  & 0.9685 & 0.0177(8)               \\
Sb3  & 6$c$  & 0          & 0          & 0.26505(6)  & 0.9646 & 0.0183(9)               \\
\multicolumn{7}{c}{CsV$_{8}$Sb$_{12}$}                                                  \\
atom & site  & $x/a$      & $y/b$      & $z/c$       & s.o.f. & $U_{\rm eq}$ (A$^{2}$)  \\
Cs   & 4$a$  & 1/2        & 0          & 1/2         & 1      & 0.0239(6)               \\
V1   & 8$i$  & 1/2        & 1/2        & 0.37349(10) & 0.9719 & 0.0041(7)               \\
V2   & 16$j$ & 1/4        & 3/4        & 0.37340(7)  & 0.9831 & 0.0068(6)               \\
V3   & 8$f$  & 1/4        & 1/4        & 1/4         & 0.9410 & 0.0089(8)               \\
Sb1  & 8$i$  & 0          & 1/2        & 0.36253(4)  & 0.9869 & 0.0077(4)               \\
Sb2  & 16$m$ & 1/2        & 0.66716(10)& 0.43449(3)  & 1      & 0.0094(4)               \\
Sb3  & 16$m$ & 1/2        & 0.33846(9) & 0.31147(3)  & 1      & 0.0078(4)               \\
Sb4  & 8$i$  & 0          & 0          & 0.22243(4)  & 1      & 0.0086(4)               \\
\hline\hline
\end{tabular}
\label{2}
\end{table}

For CsV$_{6}$Sb$_{6}$, the fit of single crystal XRD with floating site occupation factor (s.o.f) gives that the ratio of three elements is close to 1 : 6 : 5.78, possibly with a very small amount of Sb vacancies on all of three Sb sites (Table 2). But the average atomic ratios of Cs : V : Sb obtained from the EDX measurement is 1 : 6.02(7) : 6.22(8) when setting the content of Cs as 1, i.e., there are no Sb vacancies. Thus, within the error range, the CsV$_{6}$Sb$_{6}$ should be a stoichiometric compound.
In contrast, for CsV$_{8}$Sb$_{12}$, similar fit of XRD pattern shows that there are vacancies in both V and Sb sites, especially for the V on the 1D chain (V3 site in Table 2). The atomic ratio of Cs : V : Sb is 1 : 7.76 : 11.98. The EDX result also indicates that the significant V deficiencies exist in this material (Cs : V : Sb =  1 : 7.00(2) : 11.38(8) when setting the content of Cs as 1).

\begin{figure}[tbp]
\centerline{\includegraphics[scale=0.17]{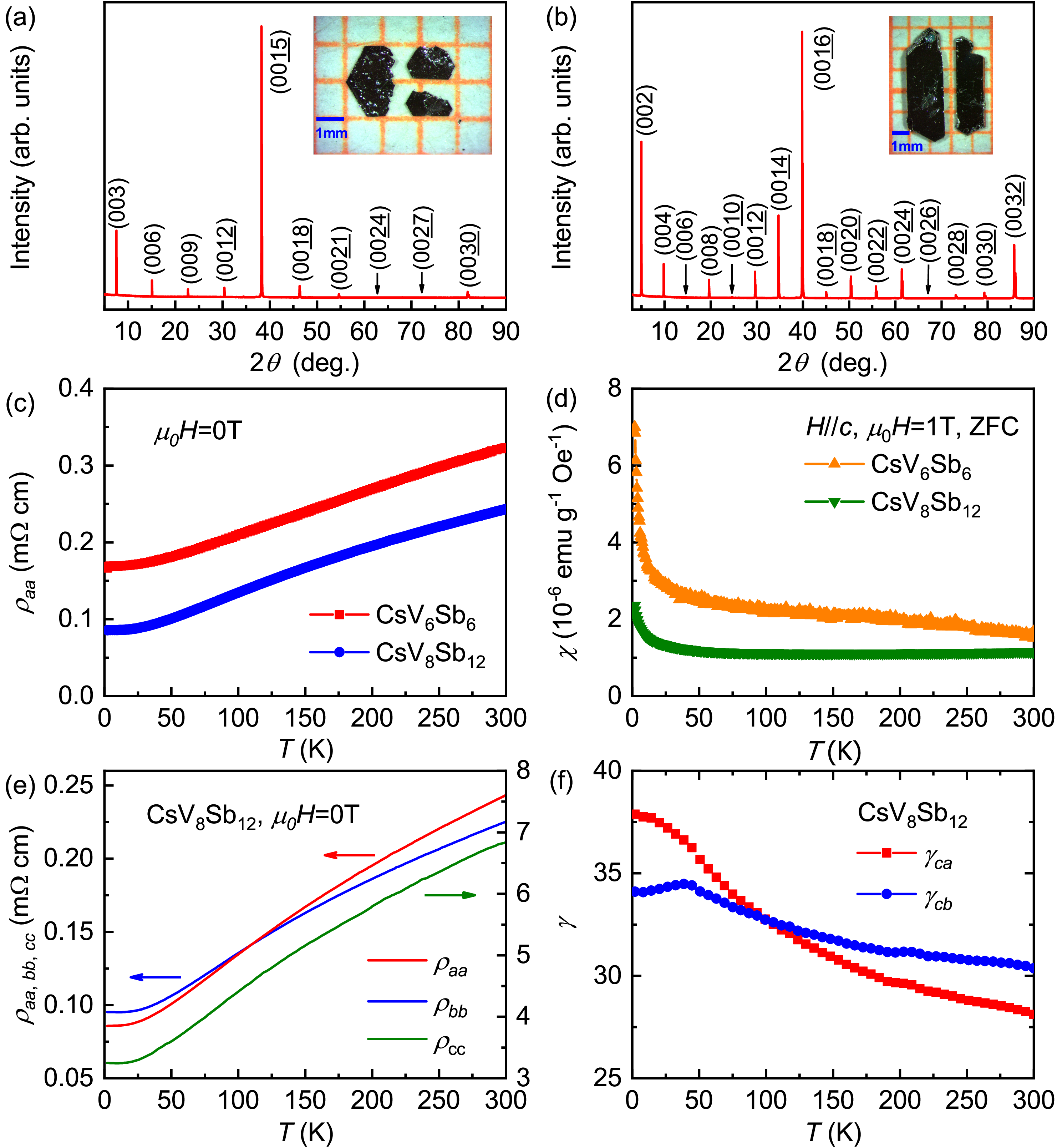}} \vspace*{-0.3cm}
\caption{(a) and (b) XRD pattern of a CsV$_{6}$Sb$_{6}$ and CsV$_{8}$Sb$_{12}$ single crystal, respectively. Insets: photos of typical CsV$_{6}$Sb$_{6}$ and CsV$_{8}$Sb$_{12}$ single crystals on a 1 mm grid paper. (c) Temperature dependence of $a$-axis resistivity $\rho_{aa}(T)$ at zero field for CsV$_{6}$Sb$_{6}$ and CsV$_{8}$Sb$_{12}$ single crystals. (d) The $\chi(T)$ as a function of temperature when $H\Vert c$ and $\mu_{0}H=$ 1 T with ZFC mode for both materials. (e) Temperature dependence of $\rho(T)$ along three crystallographic axes of CsV$_{8}$Sb$_{12}$.
(f) Temperature dependence of anisotropy of resistivity $\gamma_{ca}(T)=\rho_{cc}(T)/\rho_{aa}(T)$ and $\gamma_{cb}(T)=\rho_{cc}(T)/\rho_{bb}(T)$ for CsV$_{8}$Sb$_{12}$.}
\end{figure}

Figure 3(a) and 3(b) shows the XRD pattern of a CsV$_{6}$Sb$_{6}$ and CsV$_{8}$Sb$_{12}$ single crystal, respectively. All of peaks can be indexed by the indices of (00$l$) lattice planes. It indicates that the crystal surfaces are parallel to the $ab$-plane and perpendicular to the $c$axis for both materials.
The insets of Fig. 3(a) and (b) show the photographs of typical CsV$_{6}$Sb$_{6}$ and CsV$_{8}$Sb$_{12}$ crystals on a 1 mm grid paper. It can be seen that the shapes of these two crystals are quite different. For CsV$_{6}$Sb$_{6}$ crystals, they have a hexagonal thin-plate-like shape (inset of Fig. 3(a)), consistent with the layered structure and its rhombohedral symmetry.
In contrast, for CsV$_{8}$Sb$_{12}$ crystals, they have a rectangular shape with relatively large thickness. It reflects the orthorhombic distortion of kagome lattice.

Figure 3(c) presents temperature dependence of the zero-field $a$-axis resistivity $\rho_{aa}(T)$ for CsV$_{6}$Sb$_{6}$ and CsV$_{8}$Sb$_{12}$ single crystals.
Both of $\rho_{aa}(T)$ curves exhibit similar trend of decreasing with lowering temperature. It clearly indicates that both materials are metals.
In addition, the residual resistivity ratio, defined as $\rho_{aa}$(300 K)/$\rho_{aa}$(2 K), is about 1.94 for CsV$_{6}$Sb$_{6}$ when compared to the value of 2.8 for CsV$_{8}$Sb$_{12}$.
It has to be noted that distinctly different from AV$_{3}$Sb$_{5}$, the resistivity curves do not show any anomalies in the whole temperature range (2 K - 300 K). Moreover, as shown in Fig. 3(d), again, there are no any transitions in the magnetic susceptibility $\chi(T)$ curves at $\mu_{0}H$ = 1 T with zero-field-cooling (ZFC) mode for $H\Vert c$. These results indicate that the CDW and superconducting transitions are absent in both CsV$_{6}$Sb$_{6}$ and CsV$_{8}$Sb$_{12}$ when $T$ is above 2 K.
The $\chi(T)$ curves of both materials exhibit weak temperature-dependence at $T>$ 50 K with very small values. Such kind of Pauli paramagnetism is consistent with the itinerant features of electrons in CsV$_{6}$Sb$_{6}$ and CsV$_{8}$Sb$_{12}$ and suggests that the V ions in these materials do not have obvious local moment. The slight upturns of $\chi(T)$ curves at low temperature region could be due to the trace amount of impurities on the surface of crystals.
For CsV$_{8}$Sb$_{12}$, the out-of-plane resistivity $\rho_{cc}$ is much larger than those in-plane ones ($\rho_{aa}$ and $\rho_{bb}$) (Fig. 3(e)), which is reflected in the large anisotropy of resistivity $\gamma_{ca}=\rho_{cc}/\rho_{aa}$ and $\gamma_{cb}=\rho_{cc}/\rho_{bb}$ (Fig. 3(f)).
The values of $\gamma_{ca}$ and $\gamma_{cb}$ are about 28 and 30 at 300 K, and increase to 38 and 34 at 2 K.
Notably, these $\gamma$ values are larger than that of CsV$_{3}$Sb$_{5}$ ($\gamma\sim$ 10 at 300 K and 23 at 8 K) \cite{XiangY}, suggesting a weaker interlayer coupling in CsV$_{8}$Sb$_{12}$, i.e., the two-dimensionality is more obvious.
On the other hand, the comparable $\gamma_{ca}$ and $\gamma_{cb}$ imply the small in-plane anisotropy of CsV$_{8}$Sb$_{12}$ even it has a orthorhombic structure.
The slightly smaller $\rho_{aa}$ than $\rho_{bb}$ at low temperature could be due to the existent of V chains along the $a$-axis, in which the intrachain hopping could enhance conductivity.

\begin{figure}[tbp]
\centerline{\includegraphics[scale=0.17]{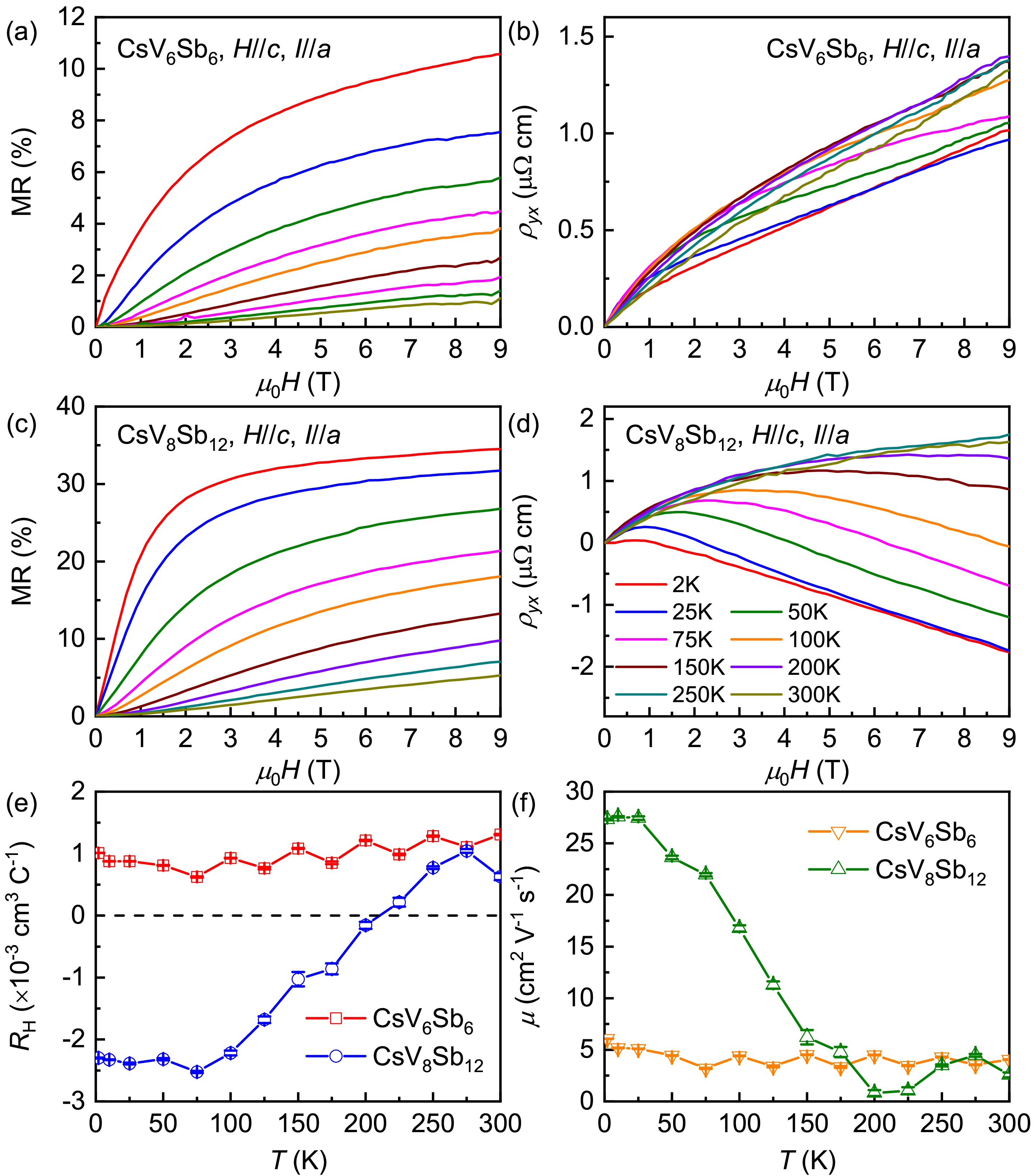}} \vspace*{-0.3cm}
\caption{Field dependence of (a, c) MR and (b, d) Hall resistivity $\rho_{yx}(\mu_{0}H)$ up to $\mu_{0}H$ = 9 T with $H\Vert c$ and $I\Vert a$ at various temperatures for CsV$_{6}$Sb$_{6}$ and CsV$_{8}$Sb$_{12}$. The color codes in (a) - (d) are same. (e) Temperature dependence of $R_{\rm H}(T)$ derived from the linear fit of $\rho_{yx}(\mu_{0}H)$ curve at high-field region. (f) Derived $\mu(T)$ as a function of temperature.}
\end{figure}

CsV$_{6}$Sb$_{6}$ and CsV$_{8}$Sb$_{12}$ exhibit similar behavior of magnetoresistance (MR $=(\rho_{xx}(T,\mu_{0}H)-\rho_{xx}(T,0))/\rho_{xx}(T,0)$) for $H\Vert c$ and $I\Vert a$. At low temperature, the MR increases with field gradually and then saturates at high field (Figs. 4(a) and 4(c)). The values at 2 K and 9 T are about 11 \% and 35 \% for CsV$_{6}$Sb$_{6}$ and CsV$_{8}$Sb$_{12}$.
These saturation behaviors of MR with relatively small high-field values are remarkably different from the unsaturated MR even at very high field with extremely large absolute values in the well-known compensated semimetals, like WTe$_{2}$ and LaBi \cite{Ali,SunS}.
In the framework of two-band model, it implies that one type of carriers has a much higher concentration than another type \cite{Pippard}.
With increasing temperature, the saturation value of MR becomes smaller and the saturation field shifts to higher field. At even higher temperatures, the saturation behavior can not be observed at $\mu_{0}H\leq$ 9 T and the MR exhibits a quasi-linear behavior.
This evolution of MR can be partially explained by the decrease of carrier mobility $\mu$ at high temperature \cite{Pippard}.

As shown in Fig. 4(b), the Hall resistivity $\rho_{yx}(\mu_{0}H)$ of CsV$_{6}$Sb$_{6}$ shows a convex shape at low-field region and changes to a nearly linear behavior at high-field region. Moreover, the slopes of $\rho_{yx}(\mu_{0}H)$ curves in the whole temperature range are positive.
In contrast, the $\rho_{yx}(\mu_{0}H)$ curves of CsV$_{8}$Sb$_{12}$ at high fields and low temperatures exhibit linear behaviors with negative slopes.
These results suggest that the dominant carriers are holes and electrons in CsV$_{6}$Sb$_{6}$ and CsV$_{8}$Sb$_{12}$, respectively.
Moreover, different from CsV$_{6}$Sb$_{6}$, the low-temperature $\rho_{yx}(\mu_{0}H)$ curve of CsV$_{8}$Sb$_{12}$ has a maximum value. It shifts to higher fields with increasing temperature and the corresponding field $H_{m}$ is larger than 9 T when $T>$ 200 K.
The Hall coefficient $R_{\rm H}(T)\equiv\rho_{yx}(\mu_{0}H)/\mu_{0}H$ determined from the linear fits of $\rho_{yx}(\mu_{0}H,T)$ curves at high-field region.
For CsV$_{6}$Sb$_{6}$, the $R_{\rm H}(T)$ shows a weak temperature dependence (Fig. 4(e)). However, the $R_{\rm H}(T)$ of CsV$_{8}$Sb$_{12}$ is almost unchanged when $T\leq$ 75 K and then increases at higher temperature with changing the sign from negative to positive at 225 K.
Because of the saturation behavior of MR, this temperature dependence and sign change of $R_{\rm H}(T)$ should not be ascribed to the type change of dominant carriers at high temperature but be explained by the shift of low-field part of $R_{\rm H}$ with $H<H_{m}$ to 9 T. For this region, carrier mobility $\mu$ has a significant influence on $R_{\rm H}$ according to the two-band model \cite{Ziman}.
At 2 K, we evaluate the apparent carrier concentration $n_{a}$ using the formula $|R_{\rm H}|=1/|e|n_{a}$ and it is 6.18(1)$\times$10$^{21}$ cm$^{-3}$ and 2.718(8)$\times$10$^{21}$ cm$^{-3}$ for CsV$_{6}$Sb$_{6}$ and CsV$_{8}$Sb$_{12}$, respectively.
In addition, according to the single-band model, the value of $\mu$ can be calculated using the formula $\mu=\sigma_{xx}(0)/|e|n_{a}\approx 1/|e|n_{a}\rho_{xx}(0)=R_{\rm H}/\rho_{xx}(0)$. The temperature dependence of derived $\mu(T)$ for both compounds is presented in Fig. 4(f).
For CsV$_{6}$Sb$_{6}$, the $\mu(T)$ is small (6.07(1) cm$^{2}$ V$^{-1}$ s$^{-1}$ at 2 K) and insensitive to temperature, implying that the dominant scattering mechanism may be the electron-impurity scattering.
In contrast, for CsV$_{8}$Sb$_{12}$, the $\mu(T)$ increases with decreasing temperature in general and reaches 27.32(8) cm$^{2}$ V$^{-1}$ s$^{-1}$ at 2 K. It suggests that the electron-phonon scattering mechanism should be the dominant and the decreased scatting rate because of the gradually frozen phonons at low temperature enhances $\mu(T)$ \cite{Ziman}.
It has to be noted that the values of $n_{a}$ and $\mu$ for both CsV$_{6}$Sb$_{6}$ and CsV$_{8}$Sb$_{12}$ are lower than those in CsV$_{3}$Sb$_{5}$ \cite{YangSY}, explaining well their much larger low-temperature resistivity than the latter \cite{YangSY}.
Moreover, the drastic changes of $n_{a}$ and $\mu$ of CsV$_{6}$Sb$_{6}$ and CsV$_{8}$Sb$_{12}$ when compared with CsV$_{3}$Sb$_{5}$ reflect their obviously different electronic structures and Fermi surfaces near Fermi energy level $E_{\rm F}$, originating from their different structures. These changes also explain the absence of CDW and superconductivity in the former materials partially because the $E_{\rm F}$ may have moved away from the van Hove singularity near the $M$ point of Brillouin zone.

\section{Conclusion}

In summary, single crystals of two new members of V-based kagome metals CsV$_{6}$Sb$_{6}$ and CsV$_{8}$Sb$_{12}$ are grown successfully. Transport and magnetization measurements indicate that both materials show metallic behaviors and Pauli paramagnetism without CDW and superconducting transitions down to 2 K.
The discovery of new V-based kagome metals proves that the kagome lattice of transition metal can be
incorporated into more complicated structures. The structural flexibility of V-based kagome metals provides a new platform to study the possible correlated topological phenomena in this kind of materials.

\section{Acknowledgments}

This work was supported by National Key R\&D Program of China (Grant No. 2018YFE0202600), Beijing Natural Science Foundation (Grant No. Z200005), National Natural Science Foundation of China (Grant No. 11822412 and 11774423), the Fundamental Research Funds for the Central Universities and Research Funds of Renmin University of China (RUC) (Grant No. 19XNLG17 and 20XNH062), the Outstanding Innovative Talents Cultivation Funded Programs 2020 of Renmin Univertity of China, Beijing National Laboratory for Condensed Matter Physics, and Collaborative Research Project of Laboratory for Materials and Structures, Institute of Innovative Research, Tokyo Institute of Technology.



$\ast$ Corresponding author: hlei@ruc.edu.cn (H. C. Lei).

\end{document}